\documentclass[11pt]{article}
\usepackage{moriond,epsfig}
\def\Journal#1#2#3#4{{#1} {\bf #2}, #3 (#4)}

\def\NPB{{\em Nucl. Phys.} B}
\def\PLB{{\em Phys. Lett.}  B}
\def\PRL{\em Phys. Rev. Lett.}

\def\be{\begin{equation}}
\def\ee{\end{equation}}
\def\bea{\begin{eqnarray}}
\def\eea{\end{eqnarray}}

\begin{document}
\vspace*{4cm}
\title{$|V_{cb}|$ and $|V_{ub}|$ measurements at Belle}
\author{ H. Ishino \\
 (for the Belle collaboration) }
\address{Department of Physics, Tokyo Institute of Technology \\
2-12-1, O-okayama, Meguro-ku, Tokyo, 152-8551 Japan}
\maketitle\abstracts{In this paper, we report on the results of 
the $|V_{cb}|$ and $|V_{ub}|$
measurements obtained from studies of inclusive and exclusive 
semileptonic decays using 30fb$^{-1}$ of $e^+ e^-$ collision data collected by 
the Belle detector at KEKB.  }
\section{Introduction}
The CKM matrix ~\cite{ckm} whose elements have to be determined through 
experiments is expressed using the Wolfenstein parameters as:
\begin{equation}
\left(
  \begin{array}{ccc}
    V_{ud} & V_{us} & V_{ub} \\
    V_{cd} & V_{cs} & V_{cb} \\
    V_{td} & V_{ts} & V_{tb} 
  \end{array}
\right)
\simeq
\left(
  \begin{array}{ccc}
    1 - \frac{\lambda^2}{2} & \lambda & \lambda^3 A (\rho-i\eta) \\
    -\lambda     & 1 - \frac{\lambda^2}{2} & \lambda^2A \\
    \lambda^3 A (1-\rho-i\eta) & -\lambda^2A & 1
  \end{array}
\right)
\label{eq:ckm}
\end{equation}
A precise determination of $|V_{cb}|$ is important since some tests of
the CKM framework is sensitive to $|V_{cb}|$; for example, the $CP$ violation
parameter $|\varepsilon_K|$ which constrains the apex position of 
the CKM unitary triangle in the $\rho - \eta$ plane is proportional to
$A^4$. 
In addition, an accurate measurement of $|V_{ub}|$ is also of 
importance because $CP$ violation
does not occur in the framework of the minimal Standard Model if
$|V_{ub}|$ is zero. 

In this paper, we present $|V_{cb}|$ and $|V_{ub}|$ measurements from
inclusive and exclusive semileptonic decays of $B$ mesons using
$e^+ e^-$ collision data collected by the Belle detector at KEKB.
The Belle detector consists of a silicon vertex detector (SVD),
a central drift chamber (CDC), aerogel cherenkov counters (ACC),
time of flight counters (TOF), electromagnetic calorimeters (ECL) and
a KLM  detector which detects $K_L$s and muons.
The SVD and the CDC perform tracking and momentum measurements 
of charged particles.
With the $dE/dx$ measurements in the CDC, the TOF and the ACC provide 
separation of pions and kaons in the momentum range up to 4GeV/$c$.
Electrons are identified with combined information from the ECL, CDC, 
ACC and TOF, and muons are identified with the KLM.
A 1.5T magnetic field is provided by a superconducting solenoid magnet
located inside the KLM.
\section{$|V_{cb}|$ measurements}
\subsection{$|V_{cb}|$ from exclusive decays}
The recent development of Heavy Quark Effective Theory (HQET) ~\cite{hqet}
provides us
an expression of the differential decay rate of $B \rightarrow D^{(*)}\ell\nu$ 
as a function of $y$, 
where $y$ is the inner product of 4-velocities of $B$ and $D^{(*)}$,
with a form factor at the zero recoil point ($y = 1$) 
whose uncertainty is theoretically
controllable and the determination of $|V_{cb}|$ with the smallest theoretical
error.

We analyzed $\bar{B}^0\rightarrow D^{*+}e^-\bar{\nu_e}$ decay
mode ~\cite{dstrlnu} with the subdecay of $D^{*+} \rightarrow D^0\pi^+$, 
$D^0 \rightarrow K^-\pi^+$ using 10.2fb$^{-1}$ data.
Since in the final state there is an undetectable neutrino,
we used a partial reconstruction method; we imposed a kinematical constraint
on missing invariant mass defined as $M_{miss} \equiv (P_{B} - P_{D^* e})^2$,
where $P_{B}$ and $P_{D^* e}$ are 4 momentum vectors of the $B$ and
the $D^{*+} e^-$ system, respectively.

We also studied the decay~\cite{dlnu} of 
$\bar{B}^0\rightarrow D^+\ell^-\bar{\nu}$, where
the $D^+$ subsequently decays into $K^+\pi^+\pi^-$.
In this decay mode, we applied a full neutrino reconstruction method;
in each event we extracted information on the neutrino from the missing momentum
($\vec{p}_{miss} = \sum_i \vec{p_i}$) and missing energy 
($E_{miss} = \sum_i E_i$), where the summation is carried out for all 
reconstructed particles, and required $M_{miss}^2 = E_{miss}^2 - 
|\vec{p}_{miss}|^2 \simeq 0$.

In both decay modes, the dominant background comes from combinatorical
background in $D^{(*)}$ reconstruction. It was estimated from the 
sideband data of the $D^{(*)}$ invariant mass distributions.

\begin{figure}
\begin{center}
\psfig{figure=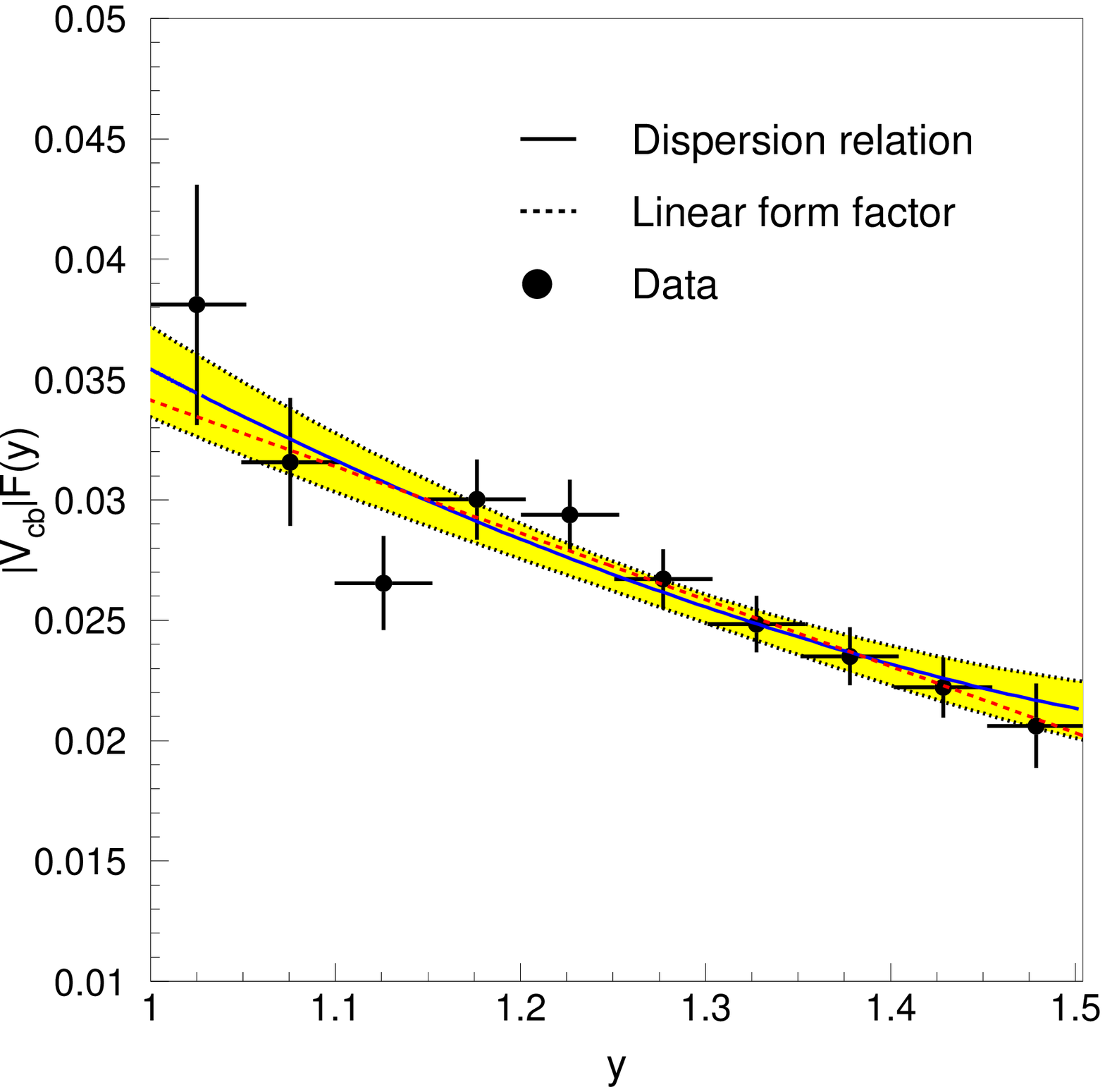,height=5cm}
\psfig{figure=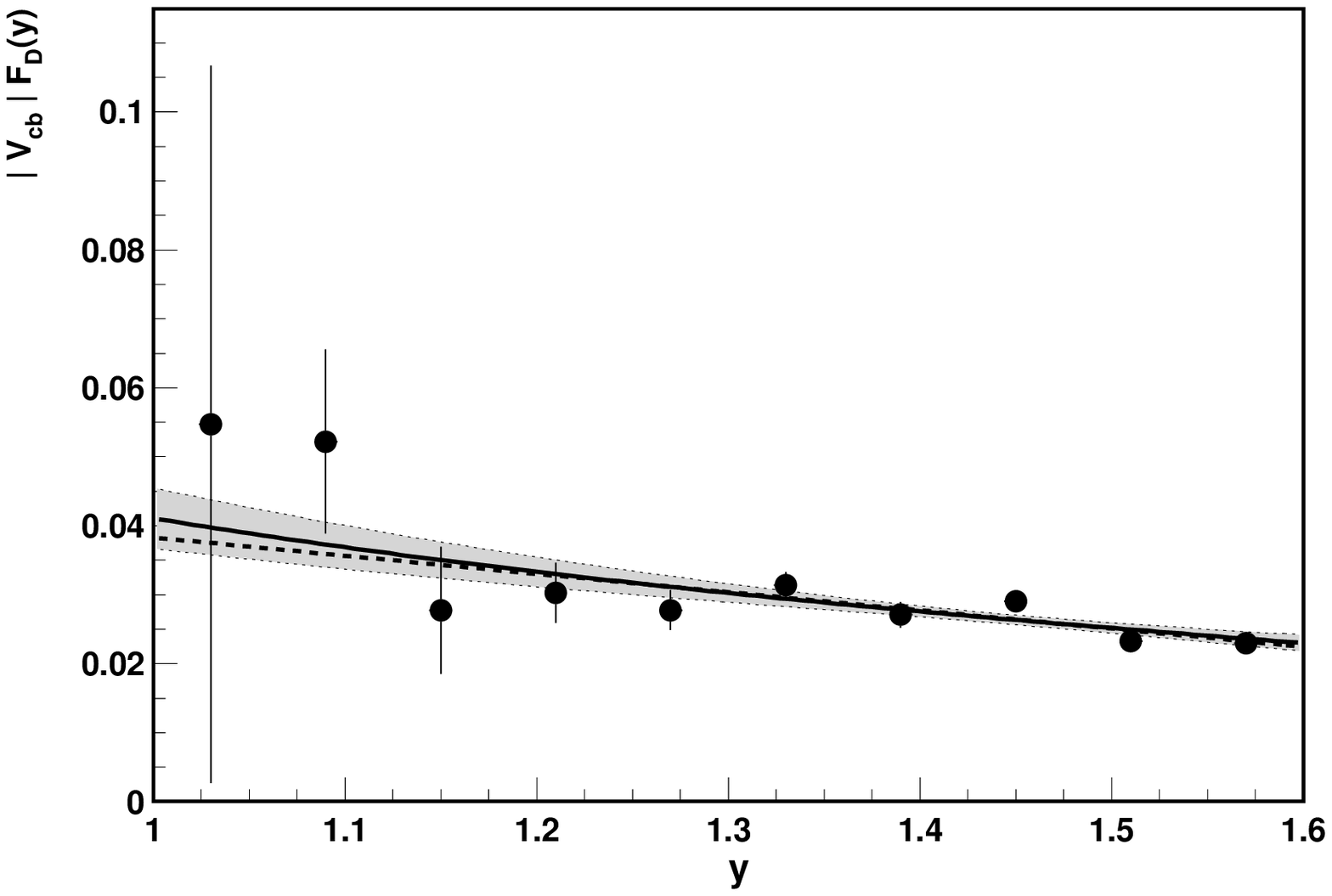,height=5cm}
\end{center}
\caption{Distributions of 
 $D^{*+}e^-\bar{\nu_e}$(left)  and $D^+\ell^-\bar{\nu}$(right) decay mode,
respectively. Fitted results are also shown.}
\label{fig:ydist}
\end{figure}

After subtracting backgrounds, correcting efficiency and unfolding
the smearing, we obtained $y$ distributions of each decay mode
as shown in figure \ref{fig:ydist}.
With the measured $y$ distributions, we fitted $|V_{cb}|F(1)$
($F(1)$ is the form factor at the zero recoil point) and $\rho^2$ which
is a single parameter determining the shape of the differential decay rate
function.
We obtained $|V_{cb}|F_{D^*}(1)=(3.54\pm0.19\pm0.18)\times10^{-2}$ and
$|V_{cb}| = (3.88\pm0.21\pm0.20\pm0.19)\times10^{-2}$ using
$F_{D^*}(1)=0.913\pm0.042$~\cite{ffdstr} 
for $\bar{B}^0\rightarrow D^{*+}e^-\bar{\nu_e}$,
$|V_{cb}|F_{D}(1)=(4.11\pm0.44\pm0.52)\times10^{-2}$ and
$|V_{cb}| = (4.19\pm0.45\pm0.53\pm0.30)\times10^{-2}$ using
$F_{D}(1)=0.98\pm0.07$~\cite{ffd} 
for $\bar{B}^0\rightarrow D^{+}\ell^-\bar{\nu}$,
where the errors are statistical, systematic and theoretical in order.
By integrating the $y$ distributions, we estimated the branching ratio
to be $Br(\bar{B}^0\rightarrow D^{*+}e^-\bar{\nu_e}) = 
(4.59\pm0.23\pm0.40)\times10^{-2}$ and
 $Br(\bar{B}^0\rightarrow D^{+}\ell^-\bar{\nu}) = 
(2.13\pm0.12\pm0.39)\times10^{-2}$.
\subsection{$|V_{cb}|$ from inclusive decays}
$|V_{cb}|$ is also obtained from the measurements of branching ratio of 
inclusive semileptonic decays using formulas based on the 
heavy quark expansion~\cite{xlnu}.
In this analysis, a high momentum lepton tag method was employed, i.e.
we tagged a high momentum lepton from the decay of one of two $B$ mesons  
and observed the lepton (the electron in this analysis) from the decay of 
the other $B$ meson.
Using the charge and kinematical correlations between the tagged lepton
and the electron, primary electrons can be distinguished from the 
secondary charm decay electrons which are the main background.
With 5.1fb$^{-1}$ data, the branching ratio was measured to be 
$Br(B\rightarrow Xe\nu) = (10.90\pm0.12(stat.)\pm0.49(syst.))\times10^{-2}$.
After subtracting the charmless semileptonic fractions
with an assumption of $Br(b\rightarrow u\ell\nu) = (0.167\pm0.055)\%$ and
using the world average $B$ lifetime of $1.607\pm0.021$ps, we
estimated $|V_{cb}|$ to be $(4.08\pm0.10\pm0.25)\times10^{-2}$, where
the first uncertainty includes statistical and systematic errors and
the second one is the theoretical error.
The $|V_{cb}|$ value obtained from the inclusive semileptonic decay analysis
is quite consistent with those from exclusive decays.
\section{$|V_{ub}|$ measurements}
\subsection{$|V_{ub}|$ from $B\rightarrow \pi\ell\nu$}
The exclusive decay $\bar{B}^0\rightarrow \pi^+\ell^-\bar{\nu}$ is 
one of the most promising modes for measuring $|V_{ub}|$.
From the branching ratio of the decay, $|V_{ub}|$ is evaluated
using formula 
$Br(\bar{B}^0\rightarrow \pi^+\ell^-\bar{\nu}) = \gamma_{\pi}|V_{ub}|^2\tau_B$,
where $\gamma_{\pi}$ is a factor determined by a model and $\tau_B$ is the
$B$ meson lifetime.
To extract the decay events, we applied the same full neutrino reconstruction 
method as employed in the $B\rightarrow D\ell\nu$ study and used 
a 29.2fb$^{-1}$ data sample.
Events which have only one lepton whose momentum ranges from 1.2 to 2.8GeV/$c$
in the $\Upsilon(4S)$ rest frame were selected.
For suppressing $B\bar{B}$ generic decay events 
which are one of the dominant backgrounds,
we also required $P_{\ell} + P_{\pi} > 3.1$GeV/$c$, since the signal $\pi$'s 
have much higher momenta than those from the background.
By fitting $\Delta E ( = E_{beam} - (E_{\pi}+E_{\ell}+E_{\nu}))$ and
lepton momentum distributions simultaneously, we obtained 770 signal events.
We adopted two models for estimating the branching ratio and $|V_{ub}|$.
The results are summarized in Table \ref{table:pilnu}.
\begin{table}[h]
\caption{Measured branching ratios (B.R.) and $|V_{ub}|$ with two models. 
The errors are statistical, systematic and theoretical
in order.}
\begin{center}
\begin{tabular}{|c|c|c|c|}
\hline
 Model & B.R. & $\gamma_{\pi}$ & $|V_{ub}|$ \\
\hline
LCSR~\cite{lcsr} & $(1.89\pm0.15\pm0.30)\times10^{-4}$ & $7.3\pm2.5$ &
                                 $(4.09\pm0.17\pm0.33\pm0.76)\times10^{-3}$ \\
UKQCD~\cite{ukqcd} & $(1.92\pm0.16\pm0.30)\times10^{-4}$ & $9^{+3}_{-2}\pm2$ &
                                 $(3.71\pm0.15\pm0.29\pm0.67)\times10^{-3}$ \\
\hline
\end{tabular}
\end{center}
\label{table:pilnu}
\end{table}
\subsection{Analysis of $B\rightarrow D_s\pi$}
\begin{figure}
\begin{center}
\psfig{figure=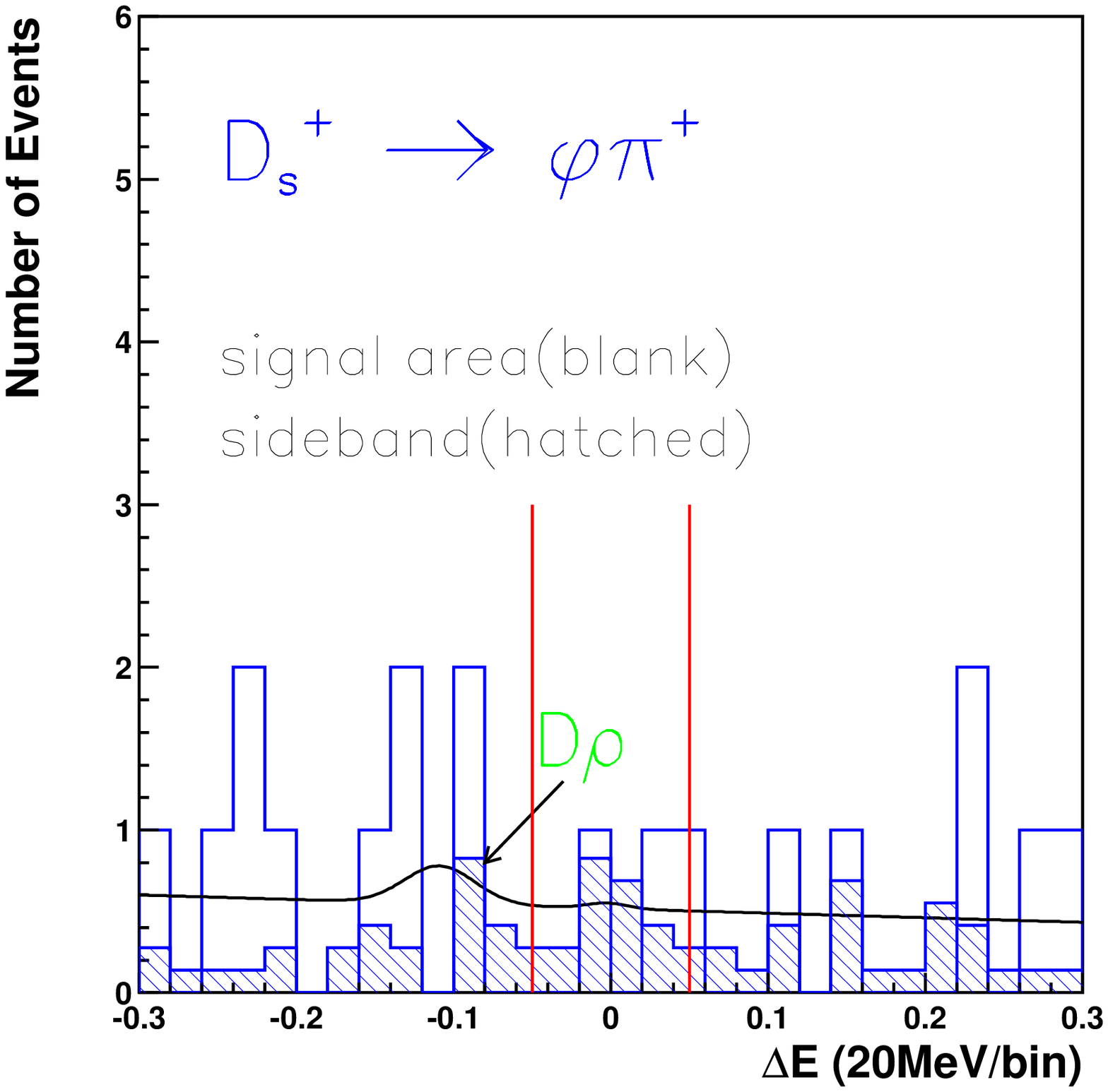,height=5cm}
\psfig{figure=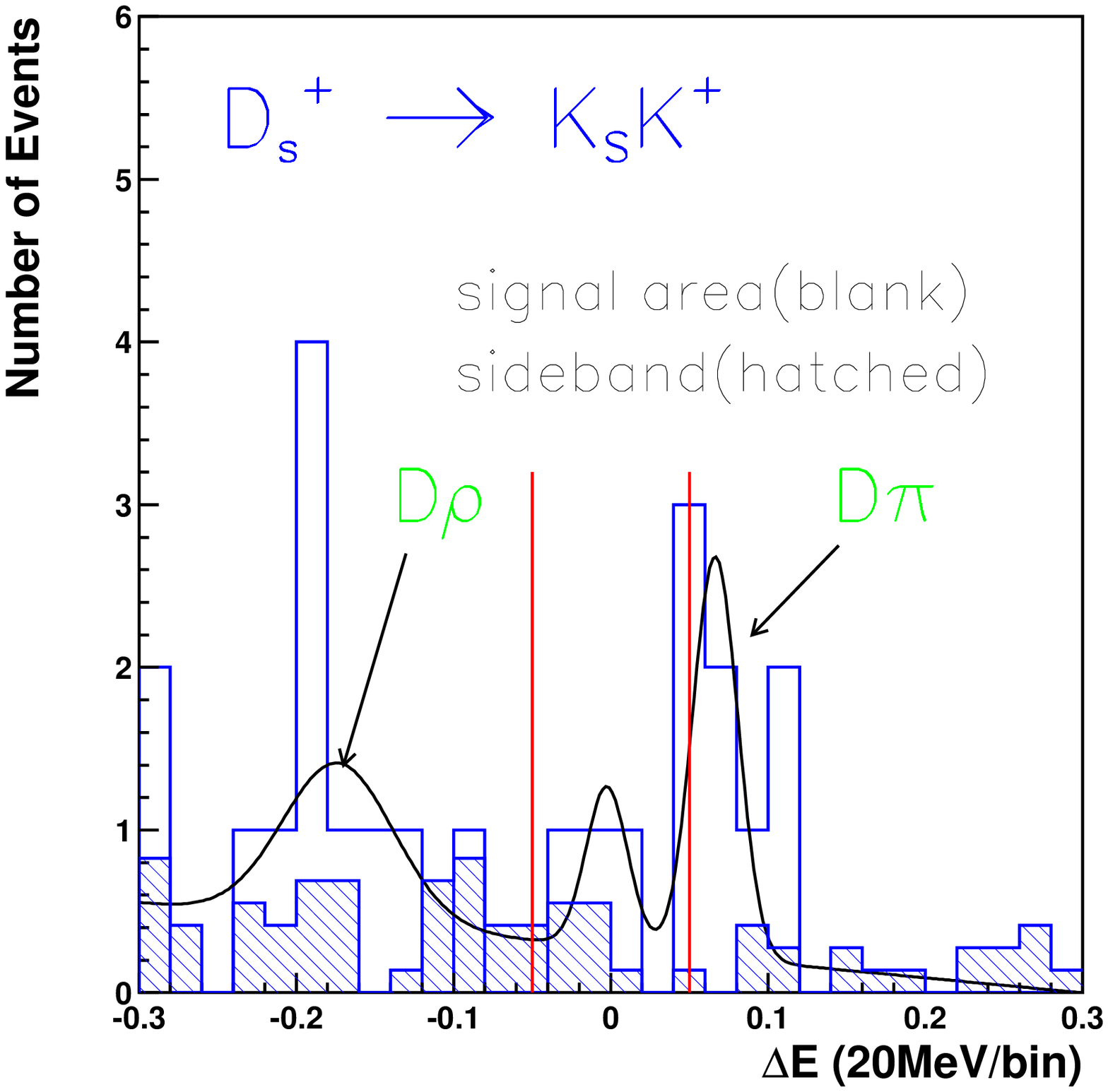,height=5cm}
\psfig{figure=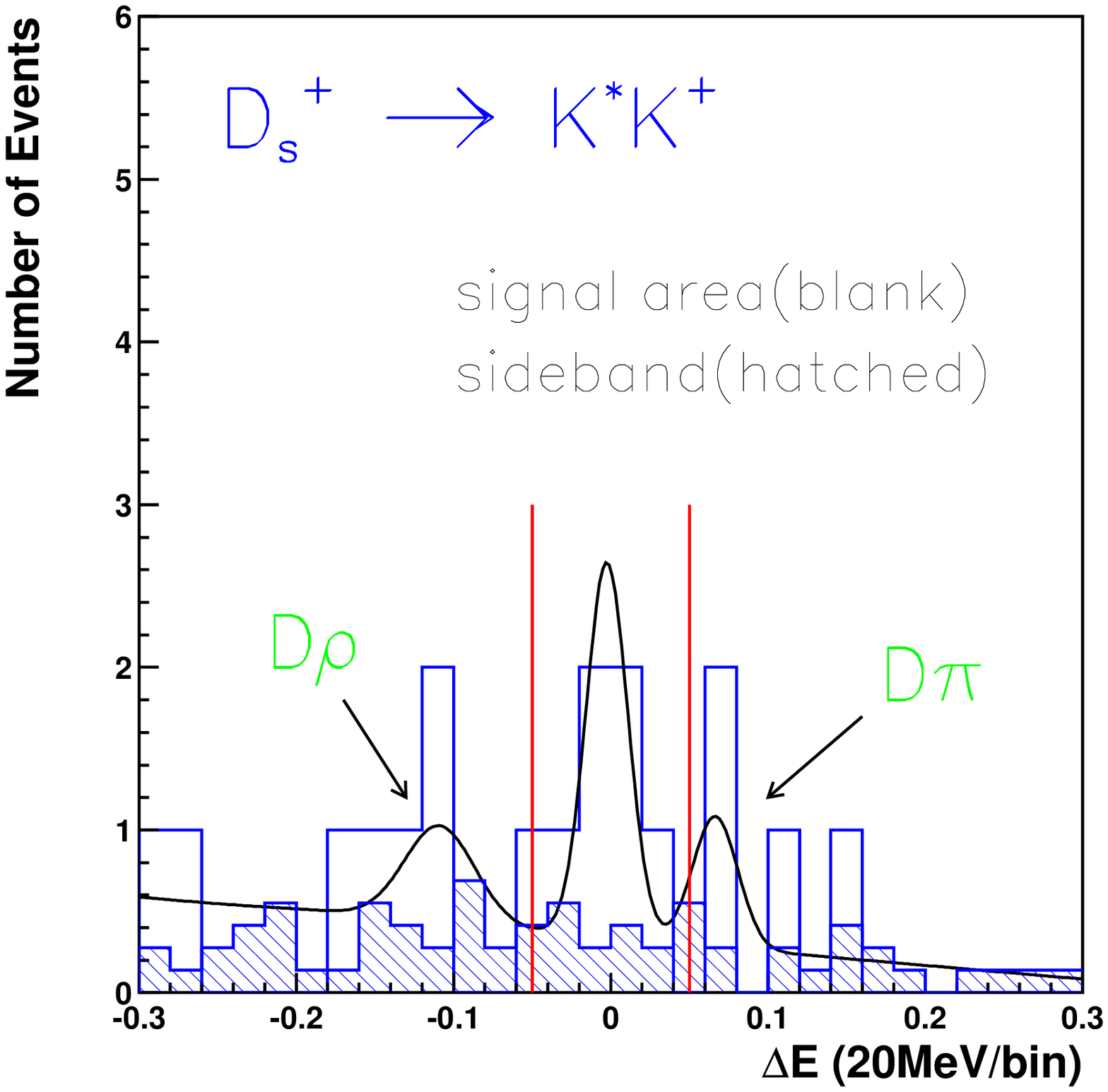,height=5cm}
\end{center}
\caption{$\Delta E$ distributions for each $D_s^+$
decay mode: from left to right, $D_s^+$ to $\phi\pi^+$, $K_sK^+$
and $K^{*0}K^+$. The blank histograms are data in the signal region,
while shaded histograms are sideband data.}
\label{fig:dspi}
\end{figure}

$|V_{ub}|$ can be extracted from the branching ratio of the hadronic decay
$B^0\rightarrow D^+_s\pi^-$ using a theoretical prediction ~\cite{dspi}
$Br(B^0\rightarrow D^+_s\pi^-) = 
(2.6\sim5.2)\times|V_{ub}^*V_{cs}|^2$.
Although the prediction has a large
uncertainty, experimental extraction of the signal events is straightforward
since signal $B$s can be fully reconstructed.
To search events, we reconstructed $D_s^+$ subsequently decaying to
$\phi(\rightarrow K^+K^-)\pi^+$, $K^0(\rightarrow \pi^+\pi^-)K^+$ and
$K^{*0}(\rightarrow K^+\pi^-)K^+$ (note that the main contribution of
systematic errors come from the uncertainties of $D_s^+$ decay
branching ratios ($25\sim30\%$)).
Figure \ref{fig:dspi} shows $\Delta E ( = E_{D_s}+E_{\pi} - E_{beam})$
distributions for each $D_s^+$ decay mode obtained using 29.2$^{-1}$ data.
The blank histograms are data in the signal region and the shaded
histograms are the sideband data.
Table \ref{table:dspi} shows reconstruction efficiencies, 
numbers of yield in the signal region and backgrounds estimated from the
sideband data and obtained upper limits at a 90\% C.L. for the three $D_s^+$
decay modes.
By combining the three modes, we obtained an upper limit
of $Br(B^0\rightarrow D^+_s\pi^-)$ to be $1.0\times10^{-4}$ at a
90\%C.L., which includes systematic errors.
\begin{table}[h]
\caption{Reconstruction efficiencies ($\varepsilon_{rec}$),
numbers of yield in the signal region and backgrounds estimated from the
sideband data and upper limits at a 90\% C.L. for the three $D_s^+$
decay modes.}
\begin{center}
\begin{tabular}{|c|c|c|c|}
\hline
 Decay mode & $\varepsilon_{rec}(\%)$ & Yield(B.G.) & B.R.($90\%$C.L.) \\
\hline
 $\phi\pi^+$ & 16.1 & 2 (2.6) &  $<0.5\times10^{-4}$ \\
 $K^0K^+$ & 12.3 & 5 (1.1) &  $<2.8\times10^{-4}$ \\
 $K^{*0}K^+$ & 8.6 & 6 (1.8) &  $<2.3\times10^{-4}$ \\
\hline
\end{tabular}
\label{table:dspi}
\end{center}
\end{table}
\section{Conclusion}
From the exclusive semileptonic decay analysis, we determined
$|V_{cb}|F_{D^*}(1) = (3.54\pm0.19\pm0.18)\times10^{-2}$ and
$|V_{cb}|F_{D}(1) = (4.11\pm0.44\pm0.52)\times10^{-2}$.
The branching ratio of inclusive semileptonic decays was measured to
be $Br(B\rightarrow X\ell\nu) = (10.90\pm0.12\pm0.49)\times10^{-2}$.
By using two models, we estimated the branching ratio of
$\bar{B}^0\rightarrow\pi^+\ell^-\bar{\nu}$ and $|V_{ub}|$ as summarized
in Table \ref{table:pilnu}.
We performed a search for $B^0\rightarrow D_s^+\pi^-$ decays and obtained
an upper limit of $1.0\times10^{-4}$ at a 90\% C.L.

\end{document}